
\font\sixrm=cmr6
\font\sixi=cmmi6
\font\sixsy=cmsy6

\font\sevenrm=cmr7
\font\seveni=cmmi7
\font\sevensy=cmsy7

\font\twelverm=cmr12
\font\twelvei=cmmi12
\font\twelvesy=cmsy10 at 12pt
\font\twelveit=cmti12
\font\twelvesl=cmsl12
\font\twelvebf=cmbx12
\font\twelvett=cmtt12

\def\twelvepoint{%
\def\rm{\fam0\twelverm}%
\def\it{\fam\itfam\twelveit}%
\def\sl{\fam\slfam\twelvesl}%
\def\bf{\fam\bffam\twelvebf}%
\def\tt{\fam\ttfam\twelvett}%
\def\cal{\twelvesy}%
 \textfont0=\twelverm
  \scriptfont0=\sevenrm
  \scriptscriptfont0=\sixrm
 \textfont1=\twelvei
  \scriptfont1=\seveni
  \scriptscriptfont1=\sixi
 \textfont2=\twelvesy
  \scriptfont2=\sevensy
  \scriptscriptfont2=\sixsy
 \textfont3=\tenex
  \scriptfont3=\tenex
  \scriptscriptfont3=\tenex
 \textfont\itfam=\twelveit
 \textfont\slfam=\twelvesl
 \textfont\bffam=\twelvebf
 \textfont\ttfam=\twelvett
 \baselineskip=15pt
}

\font\sixteenrm=cmr12 scaled\magstep1
\font\twentyrm=cmr17 scaled\magstephalf


\hsize     = 152mm
\vsize     = 215mm
\topskip   =  15pt
\parskip   =   0pt
\parindent =   0pt

\newskip\one
\one=15pt
\def\One{\vskip-\lastskip\vskip\one}

\newcount\LastMac
\def\Skipe{1}  
\def\Txe{2}    
\def\Hae{3}    
\def\Hbe{4}    

\def\SkipToFirstLine{
 \LastMac=\Skipe
 \dimen255=150pt
 \advance\dimen255 by -\pagetotal
 \vskip\dimen255
}

\def\Raggedright{%
 \rightskip=0pt plus \hsize
 \spaceskip=.3333em
 \xspaceskip=.5em
}

\def\Fullout{
 \rightskip=0pt
 \spaceskip=0pt
 \xspaceskip=0pt
}


\def\ct#1\par{
 \One
 \Raggedright
 \twentyrm\baselineskip=24pt
 #1
}

\def\ca#1\par{
 \One
 \Raggedright
 \sixteenrm\baselineskip=18pt
 \uppercase{#1}
}

\def\aa#1\par{
 \One
 \Raggedright
 \twelverm\baselineskip=15pt
 #1
}

\def\ha#1\par{
 \ifnum\LastMac=\Skipe \else \One\fi
 \LastMac=\Hae
 \Raggedright
 \twelvebf\baselineskip=15pt
 \uppercase{#1}
}

\def\hb#1\par{
 \LastMac=\Hbe
 \One
 \Raggedright
 \twelvebf\baselineskip=15pt
 #1
}

\def\tx{
 \ifnum\LastMac=\Hae \else
  \ifnum\LastMac=\Hbe \else
   \ifnum\LastMac=\Skipe \else \One
   \fi
  \fi
 \fi
 \LastMac=\Txe
 \Fullout
 \twelvepoint\rm
}

\output{\OutputPage}

\def\OutputPage{
 \shipout\vbox{\unvbox255}
}
\noindent{{\it Directions in General Relativity,} edited by B. L. Hu,
M. P. Ryan and C. V. Vishveshwara (Cambridge Univ. Press, 1993),
Vol. 1, Papers in honor of Charles Misner.}

\ct REMARKS CONCERNING THE GEOMETRIES OF
GRAVITY AND GAUGE FIELDS\par

\ca Jeeva Anandan\par

\aa Department of Physics and Astronomy,
University of South Carolina, Columbia SC 29208, USA\par

\vskip .5cm
\centerline{Abstract}\par

\tx \it {An important limitation is shown in the analogy between
the
Aharonov-Bohm effect and the parallel transport on a cone. It
illustrates a basic difference between gravity and gauge fields
due to the existence of the solder form for the space-time
geometry.
This difference is further shown by the observability of the
gravitational phase for open paths. This reinforces a previous
suggestion that the fundamental variables for quantizing the
gravitational field are the solder form and the connection, and not
the metric.}

\ha 1. Introduction\par

\tx I recall with great pleasure the discussions which I had
with Charles Misner on fundamental aspects of physics,
such as the geometry of gravity, gauge fields, and quantum
theory.
In particular, I remember the encouragement he gave to my
somewhat unorthodox attempts to understand the similarities and
differences between gauge fields and gravity from their effects on
quantum interference, and their implications to physical
geometry. It
therefore seems appropriate to present
here for his Festschrift some observations which came out of this
investigation.

\tx Geometry is a part of mathematics which can be visualized,
and is
intimately related to symmetries. This
may explain the tremendous usefulness of geometry in
physics. In section 2, I shall
make some basic remarks about the similarities and differences
between the geometries of gravity and gauge field. Then I shall
illustrate, in section 3, an important difference between them that
arises due to the existence of the solder form for gravity, using
the Aharonov-Bohm (AB) effect and parallel transport on a cone.
In
section 4, I shall further illustrate this difference by the fact that
the
gravitational phase for a spinless particle is observable for an
open
path, unlike the AB effect. This implies that the translational
gauge
symmetry of the gravitational field is broken by the existence of
the
solder form. It is then argued that the solder form and the
connection are the proper variables for quantizing the
gravitational
field.

\ha 2. Locality of Gravity and Gauge Fields\par

\tx Something which Misner emphasized to me during a
conversation was the fundamental role assigned to locality by the
theory of relativity. Already in special relativity locality is
incorporated in the fact that signals cannot travel faster than the
speed of light. But in general relativity, locality
plays an even more fundamental role: The principle of
equivalence states that the laws of physics are locally
Minkowskian.
Also, because space-time is curved, there is no distant
parallelism and vectors at two different points can only be
compared
by parallel transporting them to a common point with respect to
the
gravitational connection.

\tx These three aspects of locality are also present in gauge
fields$^1$ which
are now being used to describe the three remaining fundamental
interactions in physics. The principle of equivalence for gauge
fields
may be stated as follows: Given any point $p$ in space-time, a
gauge
can be chosen so that the corresponding connection coefficients or
vector potential vanishes at $p$ for all fields interacting with the
given gauge field. Also, there is no distant parallelism for vectors
parallel transported using the gauge field connection if the
curvature
or the Yang-Mills field strength is non vanishing.

\tx Contrary to what is sometimes said, gravity does not differ
fundamentally from gauge fields simply because it is associated
with
a metric. Because if the gauge group is unitary then it leaves
invariant
a metric in the vector space at each space-time point that consists
of
all possible values of any
matter field interacting with the gauge field at that
point. The essential difference is that the gravitational metric can
be
used to measure distances along any curve in space-time, unlike
the
gauge field metric. But I shall show now, by means of physical
arguments, that this fundmental difference between gravity and
gauge fields exists even {\it prior} to introducing the metric.

\ha 3. Aharonov-Bohm Effect and Parallel Transport on a
Cone\par

\tx It is an interesting fact that the phase shifts in quantum
interference due
to gravity and gauge fields are obtained in a simple manner from
the
distance due to the
gravitational metric and parallel transport due to gravitational
and
gauge field connections along the interfering beams$^2$.
Conversely,
the
phase shifts in quantum interference can be used to {\it define}
gauge
fields and gravity$^3$.  This is most easily shown for the simplest
gauge
field, namely the electromagnetic field, by means of the AB
effect$^4$.

\tx We recall that the magnetic AB effect is the phase shift in the
interference of two
coherent electron beams which enclose a cylinder containing a
magnetic
flux. In the interference region, the wave function may be written
as
$\psi
_1({\bf r},t) + \psi _2({\bf r},t)$, where $\psi _1$ and $\psi _2$
are
the
wave functions corresponding to the two beams. The
introduction of the magnetic field inside the cylinder modifies this
wave
function to $$\psi ({\bf r},t) = \psi _1({\bf r},t) + F_\gamma \psi
_2({\bf
r},t),$$in an appropriate gauge, where
$$F_\gamma =  exp(- {ie\over \hbar c}\oint_{\gamma} A_\mu
dx^\mu).\eqno(1)$$
Here the integral is along the curve $\gamma$ going around the
cylinder,
$A_\mu$  is the electromagnetic 4-vector potential and $e$ is the
charge of
the electron. Therefore the intensity distribution $|\psi ({\bf
r},t)|^2$
in the
interference region is modified in an apparantly non local way by
the
magnetic flux via $F_\gamma$ even though the magnetic and
electric
field strengths vanish everywhere along the beams.

\tx But this phenomenon is not surprising when we realize the
analogy
with the geometry of a cone$^5$. The cone may be formed by
taking
a flat
sheet of paper bounded by two straight lines making an angle
$\theta$ and
identifying the two straight lines (Fig. 1a); we denote this cone by
$C_\theta$. For $0 \le \theta \le 2\pi $, this is what we do when
we
make a cone by rolling this flat sheet so that these two lines
coincide
to
form one of the generators of the cone. Since the paper is not
stretched or
compressed during this process, a cone has no intrinsic curvature
except at
the apex, which can be smoothed out so that the curvature is
finite there. In the
multiply connected geometry around the apex, the intrinsic
curvature is zero everywhere, same as
the flat geometry of the
sheet which was rolled up to be the cone. In particular, a
vector is
parallel transported like on the flat sheet. But
a vector V parallel transported around a
closed curve drawn on the curvature free region of the cone so as
to
enclose the apex undergoes a
rotation by the angle $\theta$, which is the holonomy
transformation
associated with this curve.

\tx If the curvature at the apex is regarded as analogous to the
magnetic
field in the cylinder then the zero intrinsic curvature everywhere
else
corresponds to the vanishing of magnetic field strength
outside the cylinder. Then V moving in a curvature free region is
analogous to beams traveling in a field free
region. The rotation by the angle $\theta$ which relates
V and V$^\prime$ (Fig. 1a) is analogous to the phase difference
between the
two beams due to $F_\gamma$. This suggests that
the electromagnetic field may be a connection for parallel
transporting the
value of the wave function and the AB effect arises
because a
wave function when parallel transported around the closed curve
$\gamma$, gets multiplied by $F_\gamma$ . The electric and
magnetic
field strengths at each space-time
\vfil\break
\vbox{\vskip 15cm}
\centerline{Figure 1}
{\it
a) Analogy between the Aharonov-Bohm effect and parallel
transport on a
cone. The cone may be obtained by identifying the lines OA and
OA$^\prime$ on a flat sheet. Therefore, the vector V parallel
transported from B around the cone would come back to B
(identified
with B$^\prime$)
as V$^\prime$ rotated by the angle $\theta$. This is analogous to
the
AB phase shift
with the magnetic field corresponding to the curvature at the apex
of
the
cone.
b) The limitation of this analogy when $\theta$ is changed to
$\theta
+2\pi$
by adding an extra sheet of paper. The vector parallel transported
along the closed curve
BCDEB$^\prime$ rotates
by $\theta +2\pi$ with
respect to the tangent vector to the curve. This enables one to
distinguish
this cone from the earlier one. This is unlike the AB effect which
cannot
distinguish between two enclosed magnetic fluxes that differ by
one
quantum of flux.}

\tx point then constitutes the
curvature of
this connection at this point. Thus the phase factor (1) is the
holonomy
transformation associated with $\gamma$ for this connection. The
statement that the electromagnetic field is a gauge field is the
same
as
saying that it is a connection as described above.

The above mentioned conical geometry describes the
gravitational field in each section
normal to a long straight string$^6$, such as a cosmic string. A
gravitational analog of the AB effect is obtained if we interfere
two
coherent beams of identical particles with intrinsic spin around
the
string.
The resulting phase shift due to the cosmic string is a special case
of
the phase shift due to an arbitrary
gravitational field obtained before$^2$. Basically, this phase shift
consists
of two parts, one due to the change in path lengths of the
interfering
beams, and the other due to the holonomy transformation, which
in
this
case is a rotation undergone by the wave function when it is
parallel
transported around the interfering beams. This change in path
length
and
holonomy transformation, and consequently the phase shifts,
occur
even
though the space is locally flat.

\tx Now if the AB phase $$\phi _\gamma ={e\over \hbar c}
\oint_{\gamma} A_\mu
dx^\mu$$is changed by $2\pi$, which corresponds to changing the
magnetic flux inside the cylinder by a ``quantum of flux'', then (1)
is
unchanged. Therefore the AB experiment or for that matter any
other
experiment outside the cylinder cannot detect the difference
between
these two magnetic fluxes. Hence, Wu and Yang$^7$ stated that,
because of
the AB effect, the electromagnetic field strength $F_{\mu \nu}$
has
too little information,
$\phi _\gamma$ has too much information, and it is the phase
factor
or the
holonomy transformation $F_\gamma$, for arbitrary closed
curves
$\gamma$ , which has the right amount of information of the
electromagnetic field. This has been generalised to an arbitrary
connection
by the theorem$^8$ which states that from the holonomy
transformations
the connection can be reconstructed and it is then unique up to
gauge
transformations. A simple physical system to illustrate the Wu-
Yang
statement is a
superconducting ring enclosing a magnetic flux. No experiment
performed
in the interior of the ring using Cooper pairs can distinguish
between
a given enclosed flux
$\Phi$ and $\Phi +n \Phi _0$, where $\Phi _0$ is the quantum of
flux for the Cooper pair and $n$
is an integer. For example, if we measure the flux by inserting a
Josephson
junction in the ring and observe the Josephson current, we would
obtain
the same current for both fluxes. Because the AB phases for the
two
fluxes
differ by $2\pi n$ and therefore (1) is the same for both fluxes,
with
$e$ now being the charge of the Cooper pair.
\tx An important and interesting limitation of the analogy of the
AB
effect with the
cone emerges when we consider the meaning of increasing the
flux of
the curvature in the apex region of the cone by ``one
quantum''. The new flux may be regarded as corresponding to the
cone $C_{\theta
+2\pi}$ which has one extra sheet of paper compared to
$C_\theta$.
(To embed $C_{\theta +2\pi}$ into a three dimensional
Euclidean space it needs to be twisted in some way but it is well
defined by
the identification stated above.) The holonomy transformations
are
the
same for $C_\theta$ and $C_{\theta +2\pi}$ (Fig. 1b). Therefore
the
above
mentioned theorem$^8$ implies that the cones $C_\theta$ and
$C_{\theta +2\pi}$ are the same as far as their connections are
concerned. Here a connection is regarded simply
as a
rule for parallel transporting abstract vectors attached to points
on
the
cone and not regarded as tangent vectors. Physically, the phase
shift
arising from spin in an interference experiment which is
determined
by the holonomy transformation will be the same for both cones
for
a bosonic particle. For a fermionic particle, there is a difference of
$\pi$ between the phase shifts because this phase is acquired by
a
fermion when it is rotated by $2\pi$ radians. Therefore for
fermions,
$C_\theta $ is not equivalent to $C_{\theta +2\pi}$, but is
equivalent
to $C_{\theta +4\pi}$, because of the nature of the spinor
connection.
A straightforward application of the Gauss-Bonnet theorem shows
that the flux or integral
of the curvature at the smoothed out apex of the cone $C_\alpha $
is
$2\pi - \alpha$. Therefore this flux is negative when $\alpha >
2\pi$.
In the latter case, it follows
via Einstein's field equations that if $C_\alpha $ represents the
geometry around a cosmic
string then the string has negative mass. In particular, $C_{\theta
+2\pi}$ and $C_{\theta +4\pi}$ represent geomtries around cosmic
strings with negative mass.
\tx The two cones, which are the same as far as their linear
connections are
concerned, are of course, different when we take into account
their metrics. This gives rise to the phase shift due to changes in
the
path
lengths of the interfering beams$^2$. But even if we forget their
metrics,
there is a subtle difference
between the two cones. To see this, for each cone, parallel
transport a
vector around a
closed smooth curve that encloses the apex and does not intersect
itself. This vector rotates with respect to the tangent vector to the
curve by the angle $\theta$ for $C_\theta$ and by the angle
$\theta
+2\pi$ for
$C_{\theta +2\pi}$ . This difference, which can be observed by
means
of local measurements, arises ultimately because we identify
the vectors being parallel transported with the tangent vectors to
the
cone.
The mathematical concept used to make this identification in an
arbitrary
manifold is called the {\it solder form}, or the canonical 1-
form, or the canonical form$^9$. For
the electromagnetic field the vector $\psi ({\bf x},t)$ belongs to an
internal
space and cannot be compared with a tangent vector. Therefore
the
Wu-Yang statement$^7$ is valid. But the gravitational field
connection is for
parallel transporting tangent vectors. Hence there is such an
identification.
This is the most fundamental difference between gravity and
gauge
fields$^{10}$.
\tx  If $C_{\theta }$ and $C_{\theta +2\pi}$ have only the
connections or only the solder forms then they are identical. Since
the two connections in the the frame bundles over $C_{\theta }$
and
$C_{\theta +2\pi}$ have the same holonomy transformations,
there
exists a fiber bundle isomorphism $\widetilde f$ between the
frame
bundles which maps
one connection into the other$^8$. This
$\widetilde f$ induces a unique diffeomorphism $f$ between the
base manifolds $C_{\theta }$ and $C_{\theta +2\pi}$ in the
obvious
way. The differential $f_*$ is a map between tangent vectors. It
determines a fiber bundle
isomorphism $\overline f$ between the two frame bundles that
maps one solder form into the other. But $\widetilde f$ and
$\overline f$ are topologically different in the sense that one
cannot
be continuously deformed into the other. This is why we were
able to
distinguish between $C_{\theta }$ and $C_{\theta +2\pi}$ when
the connections and the solder forms are both present, even when
the metric is absent.

\ha 4. Gravitational Phase Factor\par

\tx The phase shift in quantum interference due to an arbitrary
gauge field,
which generalizes the AB effect, is determined by the ``phase
factor''{}$^{2,11}$
$$F_\gamma =  Pexp(- {ig\over \hbar c}\oint_{\gamma} A_\mu
^kT_k dx^\mu),\eqno(2)$$
where $T_k$ generate the Lie algebra of the gauge group, $A_\mu
^k$ is
the Yang-Mills gauge potential, P denotes path ordering, and
$\gamma$ is a closed curve through the interfering beams. Here,
$F_\gamma$ is an element of the gauge group. Its eigenvalues can
be
determined by interference experiments$^3$. This shows the real
significance
of (1) as an element of the $U(1)$ group, which is a special case of
the
gauge group. When $\gamma$ is an infinitesimal closed curve
spanning a
surface element represented by $d\sigma ^{\mu \nu}$,
$$F_\gamma = 1 + {i g\over 2\hbar c} F^k_{\mu \nu}T_kd\sigma
^{\mu
\nu}\eqno(3)$$
where $F^k = dA^k - g{C^k}_{ij}A^i\wedge A^j$ is the Yang-Mills
field
strength.

\tx The phase shift in quantum interference of a particle due to
the
gravitational field is determined by$^{12}$
$$F_\gamma =  Pexp[- {i\over \hbar}\int_{\gamma} (e_\mu ^a
P_a +{1\over2}{\Gamma}_\mu ^{ab} M_{ab})dx^\mu
],\eqno(4)$$
which is an element of the Poincare group that may be associated
with any path $\gamma$ in space-time. Here, $P_a$ and $M_{ab},
a,b = 0,1,2,3$ are respectively the energy-momentum and angular
momentum operators which generate the representation of the
Poincare group corresponding to the given particle, $e_\mu ^a$ is
dual to the frame $e^\mu _a$ used by local observers:
$$e_\mu ^a e^\mu _b = \delta _b^a,\eqno(5)$$
and ${\Gamma}_\mu ^{ab}$ are the connection coefficients with
respect to
this frame field. If the local observers use orthonormal frames
then
$$e^\mu _ae^\nu _bg_{\mu \nu}=\eta _{ab},\eqno(6)$$
where $g_{\mu \nu}$ is the space-time metric and $\eta _{ab}$
are
the coefficients of the Minkowski metric. When the particle has
non
zero intrinsic spin then the
values of the wave function are what are observed by observers
using the frame field $e^\mu _b$. Then (4) implies that the spinor
field is parallel transported in addition to a phase that it acquires
due to its energy-momentum. This is obtained in the WKB
approximation, disregarding here for simplicity a real factor which
does not contribute to the phase$^2$.
\tx For the special case of a spinless particle, $M_{ab}=0$, the
gravitational phase
acquired by a locally plane wave is, to a good approximation,
$$\phi = {1\over \hbar c}\int_{\gamma} e_\mu ^a p_a,\eqno(7)$$
where $p_a$ are the eigenvalues of the energy-momentum
operators
$P_a$ and the integral is along the classical trajectory$^{13}$. A
remarkable
feature of (7) is that it is observable for an
open curve $\gamma$ unlike the phase shifts for gauge fields
which
can be observed only for closed curves. For example, (7)
may be observed by the Josephson effect for a path across the
Josephson junction$^3$, or by the oscillation of strangeness in the
Kaon system for an open time-like path $\gamma$ along the Kaon
beam$^{14}$. Both these phases depend on the geometry of space-
time
as determined by the gravitational field.
\tx To understand this difference between gravity and gauge
fields
note that the field $e_\mu ^a$ plays three roles here: First,
comparing (4) with (2) suggests that $e_\mu ^a$ is like a
connection
or gauge potential associated with the translation group. Indeed,
$e_\mu ^a$ and $\Gamma _\mu ^{ab}$ may be regarded as
constituting the
connection in the affine bundle. The curvature of this connection
is
obtained by evaluating (4) for an infinitesimal closed curve
$\gamma$:
$$F_\gamma =  1+{i\over 2\hbar} (Q_{\mu \nu} ^a
P_a +{1\over 2}{R^{ab}}_{\mu \nu} M_{ab})d\sigma ^{\mu
\nu},\eqno(8)$$
using the Poincare Lie algebra, where $Q^a=de^a+{\Gamma
^a}_b\wedge e^b$
is the torsion and
$R^{ab}=d\Gamma ^{ab}+\Gamma ^a_c\wedge \Gamma ^{cb}$ is the
curvature. Eq. (8) is the
analog of (3) for
gravity. This is the most physical way that I know to regard
gravity
as the gauge field of the Poincare group.
Second, $e_\mu ^a$ represents the solder form referred to
earlier. In geometrical language, it is the pullback of the solder
form
with respect to the local section $e^\mu _b$ in the bundle of
frames,
which follows from (5).  The (Lie-algebra valued) 1-form $e_\mu
^aP_a$ acts on the tangent
vector to $\gamma$ to give an element in the Lie algebra of the
translational group, which is also an observable in the Hilbert
space.
When this observable acts on a WKB wave function it gives as an
approximate eigenvalue the
rate of change of phase along $\gamma$. When this is integrated
along $\gamma$ the phase (7) is obtained. Third, (5) and (6)
imply
that $e_\mu ^a$ is
like the square root of the metric:
$$e_\mu ^ae_\nu ^b\eta _{ab}=g_{\mu \nu}.\eqno(9)$$
\tx In the first role, there is no restriction on the values of
$e_\mu
^a$ at any given point in space-time. Indeed $e_\mu ^a$ may be
made to vanish by an appropriate choice of gauge along any
differentiable curve that does not intersect itself. The
``gravitational
field'' then has the full gauge symmetry of the affine group
$A(4,R)$,
i. e. the group of inhomogeneous linear transformations
on a four dimensional real vector space. The holonomy group
is a subgroup of the Poincare group which enables only the
generators of the Poincare Lie algebra to occur in (4). The
corresponding
``gravitational phase'', like the AB phase, would then be
meaningful
only for a closed curve $\gamma$.
\tx However, in the second role, the matrix $e_\mu ^a$ is
restricted
to be non singular. The gauge symmetry group is reduced to the
general linear group $GL(4,R)\subset A(4,R)$, with $\Gamma
_\mu
^{ab}$
being the connection or gauge field. It is the breaking of
the translational gauge symmetry that enables the phase (7) to be
observable. The solder form is the canonical 1-form defined on
the
frame bundle whose structure group is $GL(4,R)$. The discussion
of
the parallel transport of a vector around a cone in section 3 shows
the important role played by the solder form which makes this
theory richer than a gauge theory with $GL(4,R)$ as the internal
symmetry. The $e_\mu ^a$ now transforms as a tensor, instead of
a
connection, under local gauge transformations which corresponds
to
space-time dependent transformations of the frame field.
Therefore
the phase (7) is invariant under these gauge transformations for
an
open curve $\gamma$, as it should be because it is observable.

\tx Despite the breaking of the translational gauge invariance, the
torsion which appears in (8) as the curvature corresponding to
this
group nevertheless arises naturally from a physical point of view.
Because the motion of the amplitude of a spinor wave function
provides an operational definition of the connection which is
independent of the Christoffel connection that comes from the
metric$^{2,15}$.
Therefore, the connection, in a coordinate basis, can be non
symmetric and the torsion is then twice the antisymmetric part of
this connection. Hence the burden of proof is on gravitational
theories with zero torsion to justify this constraint and not
on torsion theories to justify introducing torsion, because
kinematically torsion arises naturally whenever there are fields
with
intrinsic spin as seen above. But it is not necessary to introduce a
metric in the first two roles of $e_\mu ^a$ discussed here.
\tx In the third role, the specification of the metric, which is the
same as specifying the orthonormal frame field $e^\mu _b$,
breaks
the gauge symmetry further to the Lorentz group $O(3,1,R)\subset
GL(4,R)$ that leaves this metric invariant. But from the observed
phases (7), the metric may be constructed$^{3,15}$. Therefore it
does
not
appear to be as fundamental a physical variable as the solder
form
or connection. From an operational point of view, the motion of a
quantum system in a gravitational field is influenced directly by
the
solder form and connection, and the metric seems to arise only as
a
secondary construct. Therefore in the reaction of the quantum
system on the gravitational field, which needs to be described by
quantum gravity, the solder form and the connection would be the
fundamental dynamical variables that are affected.
\tx It was therefore proposed that in quantizing the gravitational
field the variables $e_\mu ^a$ and $\Gamma _\mu ^{ab}$ should
be
quantized and not the metric$^{15}$.
The arguments in this paper which show further the important
role
played by the solder form reinforce this view. The important role
assigned to the vector potential by the AB effect finds its
counterpart
in quantum electrodynamics in which it is the vector potential
which
is quantized. Similarly, the quantum effects discussed above
which
depend on the gravitational phase factor (4) suggest that the
variables $e_\mu ^a$ and $\Gamma _\mu ^{ab}$ should be
quantized
in
order to obtain quantum gravitodynamics. It is noteworthy that
(4)
is an element of the Poincare group, even though the
curvature of space-time
classically breaks Poincare invariance. This is analogous to
the
phase factor (2) for gauge fields being an element of the
corresponding gauge group. This role of groups in the
fundamental interactions, together with the general role of
symmetry in
quantum physics, which is much more fundamental, substantive
and
determinative than
in classical physics, suggest that the way forward in physics at the
present time should perhaps be
guided by the precept 'symmetry is destiny'.
\tx {\bf Acknowledgements}

I thank P. O. Mazur, R. Penrose and R. Howard for useful remarks.
This work was partially supported by NSF grant no. PHY-8807812.

\vskip 1cm
\ha References\par

\tx \item{1.} C. N. Yang and R. L. Mills, Phys. Rev. 96 (1954) 191.
\item{2.} J. Anandan, Nuov. Cim. 53A (1979) 221.
\item{3.} J. Anandan, Phys. Rev. D 33 (1986) 2280; J. Anandan in
{\it
Topological Properties and Global Structure of Space-Time}, eds. P.
G.
Bergmann and V. De Sabbata (Plenum Press, NY 1985), p. 1-14.
\item{4.} Y. Aharonov and D. Bohm, Phys. Rev. 115 (1959) 485.
\item{5.} J. S. Dowker, Nuov. Cim. 52B (1967) 129.
\item{6.} L. Marder, Proc. Roy. Soc. A 252 (1959) 45; in {\it Recent
Devolopments in General Relativity} (Pergamon, New York 1962).
\item{7.} T. T. Wu and C. N. Yang, Phys. Rev. D, 33 (1986) 2280.
\item{8.} J. Anandan in {\it Conference on Differential Geometric
Methods
in Physics}, edited by G. Denardo and H. D. Doebner (World
Scientific,
Singapore, 1983) p. 211.
\item{9.}S. Kobayashi and K. Nomizu, {\it Foundations of
Differential
Geometry} (John Wiley, New York 1963) p. 118.
\item{10.} A. Trautman in {\it The Physicist's Conception of
Nature},
edited
by J. Mehra (Reidel, Holland, 1973).
\item{11.} See also, I. Bialynicki-Birula, Bull. Acad. Pol. Sci., Ser.
Sci.
Math. Astron. Phys. 11 (1963) 135; D. Wisnivesky and Y.
Aharonov,
Ann, of Phys. 45 (1967) 479.
\item{12.} J. Anandan in {\it Quantum Theory and Gravitation},
edited by A. R. Marlow (Academic Press, New York 1980), p. 157.
\item{13.} J. Anandan, Phys. Rev. D, 15 (1977) 1448.
\item{14.} L. Stodolsky, J. Gen. Rel. and Grav.11 (1979) 391.
\item{15.} J. Anandan, Found. Phys. 10 (1980) 601.

\end